\documentclass{article}
\usepackage{amsmath,amssymb,graphicx,amsfonts}
\usepackage{algorithmic}
\usepackage{algorithm}
\usepackage{tikz,xcolor}
\usetikzlibrary{backgrounds,decorations.pathmorphing,arrows}

\def\EE{\mathbb E}

\def\RR{\mathbb R}

\def\cI{\mathcal I}

\def\cP{\mathcal P}

\def\cN{\mathcal N}

\def\be{\mathbf e}

\def\b1{\mathbf 1}

\def\ga{\alpha}
\def\gb{\beta}

\def\gd{\delta}

\def\cI{{\mathcal I}}

\newcommand{\qed}{\hfill$\square$\bigskip}
\newcommand{\raf}[1]{(\ref{#1})}
\newcommand{\proof}{\noindent {\bf Proof}~~}

\newcommand{\poly}{\operatorname{poly}}

\newcommand{\hide}[1]{}

\newcommand{\Maxstrat}{\mathcal{K}}
\newcommand{\Minstrat}{\mathcal{L}}

\DeclareMathOperator*{\Val}{~Val~}

\newtheorem{theorem}{Theorem}

\newtheorem{lemma}{Lemma}

\newtheorem{remark}{Remark}

\newtheorem{definition}{Definition}
\newcommand{\Max}{\text{player 1}}
\newcommand{\Min}{\text{player 2}}

\title{A Potential Reduction Algorithm for Two-person Zero-sum Mean Payoff Stochastic Games
\thanks{
This research was partially supported by DIMACS, Center for Discrete Mathematics and
Theoretical Computer Science, Rutgers University, and by the Scientific Grant-in-Aid from Ministry of Education, Science, Sports and Culture of Japan. 
Part of this research was done at the Mathematisches Forschungsinstitut Oberwolfach during stays within the Research in Pairs Program. 
The first author also acknowledges the partial support of NSF Grant IIS-1161476.
}
\thanks{An extended abstract of this paper was published in the proceedings of the Combinatorial Optimization and Applications, 2014 \cite{BEGM14}.
}
}
\author{
Endre Boros\thanks{MSIS Department and RUTCOR, Rutgers University, 100 Rockafellar Road, Livingston Campus
Piscataway, NJ 08854, USA;
($\{$boros,gurvich$\}$@rutcor.rutgers.edu)}
\and
Khaled Elbassioni\thanks{Masdar Institute of Science and Technology, P.O.Box 54224, Abu Dhabi, UAE;
(kelbassioni@masdar.ac.ae)}
\and
Vladimir Gurvich\footnotemark[2]
\and
Kazuhisa Makino\thanks{Research Institute for Mathematical Sciences (RIMS)
Kyoto University, Kyoto 606-8502, Japan;
(makino@kurims.kyoto-u.ac.jp)}
}

\begin{document}
\date{}
\maketitle
\begin{abstract}
We suggest a new algorithm for
two-person zero-sum
undiscounted stochastic games
focusing on stationary strategies. Given a positive real $\epsilon$, let us call a stochastic game $\epsilon$-ergodic, if its values from any two initial positions differ by at most $\epsilon$.
The proposed new algorithm outputs for every $\epsilon>0$  in finite time either a pair of stationary strategies for the two players guaranteeing that the values from any initial positions are within an $\epsilon$-range, or identifies two initial positions $u$ and $v$ and corresponding stationary strategies for the players proving that the game values starting from $u$ and $v$ are at least $\epsilon/24$ apart.
In particular, the above result shows that if a stochastic game is $\epsilon$-ergodic, then there are stationary strategies for the players proving $24\epsilon$-ergodicity.
This result strengthens and provides a constructive version of an existential result by Vrieze (1980) claiming that if a stochastic game is $0$-ergodic, then there are $\epsilon$-optimal stationary strategies for every $\epsilon > 0$.
The suggested algorithm 
is based on a potential transformation technique
that changes the range of local values at all positions
without changing the normal form of the game.

\end{abstract}
 %%%%%%%%%%%%%%%%%%%%%%%%%%%%%%%%%%%%%%%%%%%%%%%%

{\bf keywords:} undiscounted stochastic games, 
%switching-controller,
limiting average payoff, mean payoff, local reward,
potential transformation, computational game theory% mandatory: Please provide 1-5 keywords
%%%%%%%%%%%%%%%%%%%%%%%%%%%%%%%%%%%%%%%%%%%%%%%%%%%%%%%%%
\section{Introduction}
\label{intro}

\subsection{Basic Concepts and Notation}
\label{basic}
Stochastic games were introduced in 1953 by Shapley  \cite{Sha53} for the discounted case, and extended to the undiscounted case by Gillette \cite{Gil57}.
Each such game $\Gamma =\left( p_{k\ell}^{vu},~ r_{k\ell}^{vu}~\mid~ k\in K^v,~\ell\in L^v\right.$, $\left.u,v\in V \right)$ is played by
two players on a finite set $V$ of vertices (states, or positions);
$K^v$ and $L^v$ for $v\in V$ are finite sets of actions (pure strategies) of the two players; $p_{k\ell}^{vu}\in[0,1]$ is the transition probability from state $v$ to state $u$ if players chose actions $k\in K^v$ and $\ell\in L^v$ at state $v\in V$; and
$r_{k\ell}^{vu}\in\RR$ is the reward player 1 (the maximizer) receives from player 2 (the minimizer), correpsonding to this transition. We assume that the game is non-stopping,
%\footnote{Shapley's original stochastic games were assumed to have positive \emph{stopping probabilities}, i.e., at each state $v$, $\sum_{u\in V} p_{k\ell}^{vu} <1$, and with probability $1-\sum_{u\in V} p_{k\ell}^{vu}$, the game stops at state $v$ if actions $k$ and $\ell$ are selected by the players.}
that is,
%\begin{equation*}\label{non-stopping}
$\sum_{u\in V} p_{k\ell}^{vu}=1$
%\end{equation*}
for all $v\in V$ and $k\in K^v,~\ell\in L^v$. To simplify later expressions, let us denote by $P^{vu}\in[0,1]^{K^v\times L^v}$ the transition matrix, the elements of which are the probabilities $p_{k\ell}^{vu}$, and associate in $\Gamma$ a \emph{local expected reward matrix} $A^v$ to every $v\in V$ defined by
\begin{equation}\label{reward-matrix}
(A^v)_{k\ell} ~=~ \sum_{u\in V} p_{k\ell}^{vu}r_{k\ell}^{vu}.
\end{equation}
%When the game $\Gamma$ is not clear from the context, we shall write $r_{k\ell}^{vu}(\Gamma)$, $p_{k\ell}^{vu}(\Gamma)$, $A^v(\Gamma)$, etc.

\smallskip

In the game $\Gamma$, players first agree on an initial vertex $v_0\in V$ to start. Then, in a general step $j=0,1,...$, when the game arrives to state $v_j=v\in V$, they choose mixed strategies $\alpha^v\in \Delta(K^v):=\{ y\in \mathbb{R}^{K^v} \mid \sum_{i\in K^v}y_i=1,~ y_i\geq 0 ~ \text{ for } ~ i\in K^v\}$ and $\beta^v\in \Delta(L^v)$, player 1 receives the amount of
$
b_j=\alpha^v A^v\beta^v
$
from player 2, and the game moves to the next state $u$ chosen according to the transition probabilities
%\begin{equation}\label{p-alpha-beta}
$p_{\alpha,\beta}^{vu} ~=~ \alpha^v P^{vu}\beta^v.$
%\end{equation}

%\medskip

The \emph{undiscounted limiting average (effective) payoff} is the {\it Cesaro average}
\begin{equation}\label{mean}
g^{v_0}(\Gamma) ~=~ \liminf_{N\rightarrow \infty} \frac{1}{N+1}\sum_{j=0}^{N} \mathbb{E}[b_j],
\end{equation}
where the expectation is taken over all random choices made (according to mixed strategies and transition probabilities) up to step $j$ of the play. The purpose of player 1 is to maximize $g^{v_0}(\Gamma)$, while player 2 would like to minimize it.

%Let us remark that so far we have defined the game values in terms of stationary strategies.
In 1981, Mertens and Neymann in their seminal paper
\cite{MN81} proved that
every stochastic game has a value from any initial position in terms of history dependent strategies.
An example (the so-called Big Match) showing that
the same does not hold when restricted to stationary strategies
was given in 1957 in Gillette's paper \cite{Gil57}; see also \cite{BF68}.

In this paper we shall restrict ourselves
(and the players) to the so-called
{\em stationary} strategies, that is, the mixed strategy
chosen in a position $v \in V$
can depend only on  $v$  but not on the preceding positions or moves before reaching $v$ (i.e., not on the history of the play). We will denote by $\Maxstrat(\Gamma)$ and $\Minstrat(\Gamma)$ the sets of stationary strategies of \Max\ and \Min, respectively, that is,

\begin{equation*}\label{stationary-strategies}
\Maxstrat(\Gamma) = \bigotimes_{v\in V}\Delta(K^v) ~~~\text{ and }~~~ \Minstrat(\Gamma) = \bigotimes_{v\in V}\Delta(L^v).
\end{equation*}

\noindent Vrieze (1980) showed that if a stochastic game $\Gamma$ has a value $g^{v_0}(\Gamma)=m$, which is a constant, independent of the initial state $v_0\in V$, then it has a value in $\epsilon$-optimal stationary strategies for any $\epsilon>0$. We call such games {\it ergodic} and extend their definition as follows.

\begin{definition}\label{d1}
For $\epsilon>0$, a stochastic game $\Gamma$ is said to be $\epsilon$-\emph{ergodic} if the game values from any two initial positions differ by at most $\epsilon$, that is, $|g^v(\Gamma)-g^u(\Gamma)|\le\epsilon$, for all $u,v\in V$. A $0$-ergodic game will be simply called ergodic.
\end{definition}

Our main result in this paper is an algorithm that decides, for any given stochastic game $\Gamma$ and $\epsilon>0$, whether or not $\Gamma$ is $\epsilon$-ergodic, and provides a witness for its $\epsilon$-ergodicity/non-ergodicity. As a corollary, we get a constructive proof of the above mentioned theorem of Vrieze \cite{V80}. A notion central to our algorithm is the concept of a {\it potential transformation} introduced in the following section.

\hide{
\begin{remark}
\label{Abel-Cesaro}
Let us mention the so-called \emph{discounted limit payoff}
defined as the Abel average

\[
g^w(\gd) =  ~=~ (1 - \gd)\lim_{N \rightarrow \infty} \sum_{j=0}^N \gd^j \EE[b_j]
\]

\noindent
which is also common in the literature.
The parameter $\gd\in(0, 1)$
is interpreted as a discount factor.
Remarkably,  $g^w(\gd) \rightarrow g^w$ as $\gd \rightarrow 1$,
for the sequences $\{\EE[b_i]\}_{i=0}^\infty$,
related to the Markov chains, controlled processes, and
stochastic games \cite{Gil57, Bla62, LL69}.

Potential transformations were defined
for the $\gd$-discounted games
in \cite{GKK88} by a slightly modified formula:
$r^{v,u}_{k\ell}(x) = r_{k\ell}^{vu} + x^v - \gd x^u.$
It is immediate to verify that
such a transformation results in adding
the same constant $(1 - \gd) x^{v_0}$
to the Abel average values of all plays beginning in  $v_0$.

The undiscounted games with limiting average (or mean) payoff
were introduced in 1957 by Gillette in his seminal paper
\cite{Gil57}.
These games appear to be
more difficult for analysis than the discounted ones, because
contracting mappings can be efficiently applied in the latter case.
For this reason, most known algorithms are developed for the discounted games and
then adapted for the undiscounted case by means of the above limit transition.
The strategy iteration algorithm
of \cite{HK66} given by Hoffman and Karp in 1966 is an exception;
another exception is the algorithm suggested in the present paper.
\end{remark}
}

\subsection{Potential transformations}

In 1958 Gallai \cite{Gal58} suggested the following
simple transformation.
Let  $x : V \rightarrow \RR$  be a mapping
that assigns to each state  $v \in V$  a real number  $x^v$
called the {\em potential} of  $v$.
For every transition  $(v, u)$ and pair of actions $k\in K^v$ and $\ell\in L^v$ let us transform the payoff
$r_{k\ell}^{vu}$ as follows:
\[
r_{k\ell}^{vu}(x) = r_{k\ell}^{vu} + x^v - x^u.
\]
Then the one step expected payoff amount changes to $\EE[b_j(x)] = \EE[b_j] + \EE[x^{v_{j}}] - \EE[x^{v_{j+1}}]$, where $v_j\in V$ is the (random) position reached at step $j$ of the play. However, as the sum of these expectations telescopes, the limiting average payoff remains the same for all finite potentials:
\[
g^{v_0}(\Gamma(x)) = g^{v_0}(\Gamma) + \lim_{N\rightarrow\infty} \frac{1}{N}\EE[x^{v_0}-x^{v_N}] = g^{v_0}(\Gamma).
\]
Thus, the transformed game remains equivalent with
the original one.

Using potential transformations we may be able to obtain a proof for ergodicity/non-ergodicity. This is made more precise in the following section.
%\begin{definition}\label{d2}
%For $\epsilon>0$, a stochastic game $\Gamma$ is said to be  {\it locally} $\epsilon$-\emph{ergodic} if there is a potential transformation $x\in\RR^V$ such that $|m^v(x)-m^u(x)|\leq \epsilon$ for any two states  $u, v \in V$. 
%\end{definition}
%
%Let us add, justifying the above definition, that if for some potential transformation we have $m^v(x)\in [a,b]$ for all $v\in V$, for some $a<b$, then the game's values also belong to $[a,b]$ from any initial position \cite{BEGM-DGAA}.

$m^v$ is the value of the matrix game $A^v$ at state $v$.
\subsection{Local and Global Values and Concepts of Ergodicity}
\label{localglobal}
Let us consider an arbitrary potential $x\in \mathbb{R}^V$, and define the {\em local value} $m^v(x)$  at position  $v \in V$ as the value of the $|K^v| \times |L^v|$ local reward matrix game $A^v(x)$  with entries
\begin{equation}\label{e99}
a^v_{k\ell}(x) = \sum_{u \in V} p_{k \ell}^{v u} (r_{k \ell}^{v u}+x^v-x^u), ~~~\text{ for all } k \in K^v, \ell \in L^v,
\end{equation}
that is,
{\small
\[
m^v(x) ~=~
\Val(A^v(x)):=\max_{\alpha^v\in\Delta(K^v)}~\min_{\beta^v\in\Delta(L^v)} ~\ga^v A^v(x) \gb^v =
\min_{\beta^v\in\Delta(L^v)}~\max_{\alpha^v\in\Delta(K^v)}\ga^v A^v(x) \gb^v.
\]
}
To a pair of stationary strategies $\alpha =(\alpha^v|v\in V)\in \Maxstrat(\Gamma)$ and $\beta =(\beta^v|v\in V)\in \Minstrat(\Gamma)$
we associate a Markov chain $\mathcal{M}_{\alpha,\beta}(\Gamma)$ on states in $V$, defined by the transition probabilities $p_{\alpha,\beta}^{vu} ~=~ \alpha^v P^{vu}\beta^v$. Then, this Markov chain has unique limiting probability distributions $(q^{vu}_{\alpha,\beta}|\ u\in V)$, where $q^{vu}_{\alpha,\beta}$ is the probability of staying in state $u\in V$ when the initial vertex is $v\in V$. With this notation, The limiting average payoff \raf{mean} starting from vertex $v\in V$ can be computed as
\begin{equation}\label{g-v-alpha-beta}
g^v(\alpha,\beta) ~=~ \sum_{u\in V}q_{\alpha,\beta}^{vu}\,\left(\alpha^uA^u\beta^u\right).
\end{equation}
The game is said to be to be solvable in {\it uniformly optimal} stationary strategies, if there exist stationary strategies $\bar\alpha \in \Maxstrat(\Gamma)$ and $\bar\beta \in \Minstrat(\Gamma)$, such that for all initial states $v\in V$
\begin{equation}\label{e-opt}
g^v(\bar\alpha,\bar\beta) ~=~ \max_{\alpha\in \Maxstrat(\Gamma)}g^v(\alpha,\bar\beta) ~=~ \min_{\beta\in \Minstrat(\Gamma)}g^v(\bar\alpha,\beta).
\end{equation}
This common quantity, if exists, is the value of the game with initial position $v\in V$, and will be simply denoted by $g^v=g^v(\Gamma)$.

\subsection{Main Result}
Given an undiscounted zero-sum stochastic game, we try to reduce
the range of its local values
by a potential transformation  $x\in \RR^V$.
If they are
equalized by some potential  $x$, that is,  $m^v(x) = m$
is a constant for all  $v \in V$, we say that
the game is brought to its {\it ergodic canonical form} \cite{DGAA13}.
In this case, one can show that 
the values  $g^v$  exist and are equal to  $m$
for all initial positions  $v \in V$,
and furthermore, locally optimal strategies are globally optimal \cite{DGAA13}.
Thus, the game is solved in
uniformly optimal strategies.
However, typically we are not that lucky.

%We will give a constructive proof of the following result.
%\begin{theorem}\label{t2}
%If a stochastic game is locally $\epsilon$-ergodic then it is also $\epsilon$-ergodic. Conversely, if it is $\epsilon$-ergodic, then it is also locally $24\epsilon$-ergodic.
%\end{theorem}
%
%This follows from our main result, Theorem \ref{t-main}.
%Let us also add that local $\epsilon$-ergodicity in this paper will always be guaranteed by stationary strategies.

To state our main theorem, we need more notation.
\begin{itemize}
\item[$\bullet$]  $W>0$ is smallest integer s.t. either $p_{k\ell}^{vu}=0$ or $p_{k\ell}^{vu}\geq 1/W$
\item[$\bullet$] $R$ is the smallest real s.t.
\begin{equation}\label{e77}
0\leq r_{k\ell}^{vu}\leq R
\end{equation}
\item[$\bullet$] $N=\max_{v\in V}\{\max\{|K^v|,|L^v|\}\}$. 
\item[$\bullet$] $n=|V|$
\item[$\bullet$] $\eta=\max\{\log_2 R,\log_2 W\}$ (maximum "bit length")
\end{itemize}

{\large
\begin{theorem}\label{t-main}
For every stochastic game and $\epsilon >0$ we can find in $\left(\frac{n NWR}{\epsilon}\right)^{O(2^{2n}nN)}$ time either a potential vector $x\in \mathbb{R}^V$ proving that the game is $(24\epsilon)$-ergodic, or stationary strategies for the players proving that it is not $\epsilon$-ergodic.
\end{theorem}
}
The proof of Theorem~\ref{t-main} will be given in Section~\ref{sg}. One major hurdle that we face is that the range of potentials can grow doubly exponentially as iterations proceed, leading to much worse bounds than those stated in the theorem. To deal with this issue, we use quantifier elimination techniques \cite{BPR96,GV88,R92} to reduce the range of potentials after each iteration; see the discussion preceding Lemma~\ref{t1}.  %Although the bound given by Theorem~\ref{t-main} is doubly exponential in general, it is {\it pseudo-polynomial} for a constant number of states. Furthermore, the analysis can be 
%exponential in some interesting special cases such as the so-called
%{\em switching-controller} games and
%pseudo-polynomial for the subcase when
%all but a fixed number of positions
%are fully controlled by only one of the two players player.

\section{Related Work}

The above definition of ergodicity
follows Moulin's concept of the ergodic
extension of a matrix game \cite{Mou76}
(which is a very special example of
a stochastic game with perfect information).
Let us note that
slightly different terminology is used
in the Markov chain theory; see, for example, \cite{KS63}.

The following four algorithms
for undiscounted stochastic games are based
on stronger "ergodicity type"  conditions:
the strategy iteration algorithm
by Hoffman and Karp \cite{HK66} requires that
for any pair of stationary strategies of the two players
the obtained Markov chain has to be irreducible;
two value iteration algorithms by Federgruen
are based on similar but slightly weaker requirements;
see \cite{F80}
%page 795
for the definitions and more details; the recent algorithm of Chatterjee and
Ibsen-Jensen \cite{CI14} assumes a weaker requirement than the strong ergodicity required by Hoffman and Karp \cite{HK66}: 
they call  a stochastic game {\it almost surely ergodic} if for any pair of (not necessarily stationary) strategies of the two players, and any starting position, some strongly ergodic class (in the sense of \cite{HK66}) is reached with probability 1.

While these restrictions apply to the structure of the game, our ergodicity definition only restricts the value. Moreover,  the results in \cite{HK66} and \cite{CI14} apply to a game that already satisfies the ergodicity assumption, which seems to be hard to check. Our algorithm, on the other hand, always produces an answer, regardless whether the game is ergodic or not.  

\smallskip

Interestingly, potentials appear in \cite{F80} implicitly, as
the differences of local values of positions, as well as
in \cite{HK66}, as the dual variables
to linear programs corresponding to the controlled
Markov processes, which appear when
a player optimizes his strategy against a given strategy of the opponent.
Yet, the potential transformation is not considered
explicitly in these papers.

We prove Theorem~\ref{t-main} by an algorithm that
extends the approach
recently obtained for ergodic stochastic games
with perfect information \cite{IPCO-2010} and extended to the general (not necessarily ergodic) case in \cite{BEGM-ICALP13}.
This approach is also somewhat similar
to the first of two value iteration algorithms
suggested by Federgruen in \cite{F80}, though our approach has some distinct characteristics:
It is assumed in \cite{F80} that
the values  $g^v$  exist
and are equal for all $v$;
in particular, this assumption implies
the  $\epsilon$-ergodicity  for every  $\epsilon > 0$.
For our approach we do not need such an assumption. We can verify $\epsilon$-ergodicity for
an arbitrary given $\epsilon > 0$, or provide a proof for non-ergodicity (with a small gap) in a finite time.
Moreover,  while the approach of \cite{F80} was only shown to converge, we provide a bound in terms of the input parameters for the number of steps.

Several other algorithms for solving undiscounted zero-sum stochastic games
in stationary strategies are surveyed by
Raghavan and Filar; see Sections 4 (B) and 5 in \cite{RF91}. The only algorithmic results that we are aware of that provide bounds on the running time for approximating the value of general (undiscounted) stochastic games are those given in \cite{CMH08,HKLMT11}: in \cite{CMH08}, the authors provide an algorithm that approximates, within any factor of $\epsilon>0$,  the value of any stochastic game (in history dependent strategies) in time $(nN)^{nN}\poly(\eta,\log\frac{1}{\epsilon})$. In \cite{HKLMT11}, the authors give algorithms for discounted and recursive stochastic games that run in time $2^{N^{O(N^2)}} \poly(\eta,\log (\frac{1}{\epsilon}))$, and claim also that similar bounds can be obtained for general stochastic games, by reducing them to the discounted version using a discount factor of $\gd=\epsilon^{\eta N^{O(n^2)}}$ (and this bound on $\gd$ is almost tight \cite{M11}). These results are based on quantifier elimination techniques and yield very complicated history-dependent strategies. 
%Thus, while Theorem \ref{t-main} provides only an approximation scheme (the running time is polynomial in $\frac{1}{\epsilon})$ rather than $\log \frac{1}{\epsilon})$, it depends only polynomially on $N$ (similar to \cite{HKLMT11}, but unlike \cite{CMH08}). 
For almost sure ergodic games, a variant of the algorithm of Hoffman and Karp \cite{HK66} was given in \cite{CI14}; this algorithm finds $\epsilon$-optimal stationary strategies in time (roughly) $\left(\frac{Nn^2W^n}{\epsilon}\right)^{nN}\poly(N,\eta)$. 
This result is not comparable to ours, since the class of games they deal with are somewhat different (although both generalize the class of strongly ergodic games of \cite{HK66}).
Furthermore, the algorithm in Theorem \ref{t-main} exhibits the additional feature that it either provides a solution in stationary strategies in the ergodic case, if one exists, or produces a pair of stationary strategies that witness the non-ergodicity. 

\section{Pumping Algorithm}
\label{pump}

We begin by describing our procedure on an abstract level. Then we specialize it to stochastic games in Section~\ref{sg}.

Given a subset $S\subseteq V$, let us denote by $e_S\in \{0,1\}^V$ the characteristic vector of $S$.

Let us further assume that $m^v(x)$ for $v\in V$ are functions depending on \emph{potentials} $x\in \RR^n$ (where $n=|V|$) and satisfying the following properties for all subsets $S\subseteq V$ and reals $\delta\geq 0$:

\begin{itemize}
\item[(i)] $m^v(x-\delta e_S)$ is a monotone decreasing function of $\delta$ if $v\in S$;
\item[(ii)] $m^v(x-\delta e_S)$ is a monotone increasing function of $\delta$ if $v\not\in S$;
\item[(iii)] $|m^v(x)-m^v(x-\delta e_S)|\leq \delta$ for all $v\in V$.
\end{itemize}

We show in this section that under the above conditions we can change iteratively the potentials to some $x'\in\RR^n$ such that either all values $m^v(x')$, $v\in V$, are very close to one another or we can find a decomposition of the states $V$ into disjoint subsets proving that such convergence of the values is not possible.

Our main procedure is described in Algorithm~\ref{algo1} below. Given the current vector of potentials $x_\tau$ at iteration $\tau$, the procedure partitions the set of vertices into four sets according to the local value $m^v(x)$. If either the first (top) set $T_\tau$ or forth (bottom) set $B_\tau$ is empty, the procedure terminates; otherwise, the potentials of all the vertices in the first and second sets are reduced by the same amount $\delta$, and the computation proceeds to the next iteration.   

\begin{algorithm}
\caption{\textsc{Pump$(x,S)$}}
\label{algo1}
\begin{algorithmic}[1]
\REQUIRE a stochastic game $\Gamma$ a subset $S$ of states.
\ENSURE a potential $x\in\RR^S$.
\STATE Initialize $\tau:=0$, and $x_\tau:=x$. \label{(P0)}
\STATE Set $m^+:=\max_{v\in S} m^v(x_\tau)$, ~$m^-:=\min_{v\in S} m^v(x_\tau)$, and $\delta:=(m^+ -m^-)/4$.\label{(P1)}
\STATE Define
\[
\begin{array}{rl}
T_\tau &:= \{v\in S\mid m^v(x_\tau)\geq m^-+3 \delta\}\\*[3mm]
B_\tau &:= \{v\in S\mid m^v(x_\tau)< m^-+ \delta\}\\*[3mm]
M_\tau &:= S\setminus (T_\tau\cup B_\tau).
\end{array}
\] \label{(P2)}
\IF {$T_\tau=\emptyset$ or $B_\tau=\emptyset$}
   \RETURN $x_\tau$ \label{(P3)}
\ENDIF
\STATE Otherwise, set $P_\tau:=\{v\in S\mid m^v(x_\tau)\geq m^-+2\delta\}$ and update
\[
\begin{array}{rl}
x_{\tau+1}^v:=\left\{\begin{array}{ll}
x_\tau^v-\delta&\text{if } v\in P_\tau\\
x_\tau^v&\text{otherwise}.
\end{array}\right.
\end{array}
\]\label{(P4)}
\STATE Set $\tau:=\tau+1$ and Goto step \ref{(P2)}.\label{(P5)}
 \end{algorithmic}
\end{algorithm}

We can show next that properties (i), (ii) and (iii) above guarantee some simple properties for the above procedure.

\begin{lemma}\label{l1}
We have $T_{\tau+1}\subseteq T_\tau$, $B_{\tau+1}\subseteq B_\tau$ and $M_{\tau+1}\supseteq M_{\tau}$ for all iterations $\tau=0,1,\ldots$
\end{lemma}
\proof
Indeed, by (i) and (iii) we can conclude that $m^v(x_\tau)\geq m^-+\delta$ holds for all $v\in P_\tau$. Analogously, by (ii) and (iii) $m^v(x_\tau)< m^-+3\delta$ follows for all $v\not\in P_\tau$.
\qed

\begin{lemma}\label{l2}
Either $T_\tau =\emptyset$ or $B_\tau=\emptyset$ for some finite $\tau$, or there are nonempty disjoint subsets $I,F\subseteq S$, $I\supseteq T_\tau$, $F\supseteq B_\tau$, and a threshold $\tau_0$, such that for every real $\Delta\ge 0$ there exists a finite index $\tau(\Delta)\geq \tau_0$ such that
\begin{itemize}
\item[(a)] $m^v(x_\tau)\geq m^-+2\delta$ for all $v\in I$ and $m^v(x_\tau)<m^-+2\delta$ for all $v\in F$, and for all $\tau\geq \tau_0$;
\item[(b)] $x_\tau^u-x_\tau^v\geq \Delta$ for all $v\in I$ and $u\not\in I$, and for all $\tau\geq\tau(\Delta)$;
\item[(c)] $x_\tau^v-x_\tau^u\geq \Delta$ for all $v\in F$ and $u\not\in F$, and for all $\tau\geq\tau(\Delta)$.
\end{itemize}
\end{lemma}
\proof
By Lemma \ref{l1} sets $T_\tau$ and $B_\tau$ can change only monotonically, and hence only at most $|S|$ times. Thus, if \textsc{Pump}$(x,S)$ does not stop in a finite number of iterations, then after a finite number of iterations the sets $T_\tau$ and $B_\tau$ will never change and all positions in $T_\tau$ remain always pumped (that is, have their potentials reduced), while all positions in $B_\tau$ will be never pumped again.

Assuming now that the pumping algorithm \textsc{Pump}$(x,S)$ does not terminate, let us define the subset $I\subseteq S$ as the set of all those positions which are always pumped with the exception of a finite number of iterations. Analogously, let $F$ be the subset of all those positions that are never pumped with the exception of a finite number of iterations. 
Since $I$ and $F$ are finite sets, there must exist a finite $\tau_0$ such that for all $\tau\geq \tau_0$ we have $I\subseteq P_\tau$ and $F\cap P_\tau=\emptyset$, implying (a). Note that any vertex in $T_\tau$ is always pumped by (iii) and hence $T_\tau\subseteq I$ for any $\tau\ge \tau_0$; similarly, $B_\tau\subseteq F$ for any $\tau\ge \tau_0$.

Let us next observe that all positions not in $I\cup F$ are both pumped and not pumped infinitely many times. Thus, since $\delta$ is a fixed constant, for every $\Delta$ there must exist an iteration $\tau(\Delta)\geq \tau_0$ such that all
positions not in $I$ are not pumped by at least $\Delta/\delta$
many more times than those in $I$, and all positions not in $F$ are pumped by at least $\Delta/\delta$ many more times than those in $F$, implying (b) and (c).
\qed

%\bigskip

Let us next describe the use of \textsc{Pump}$(x,S)$ for repeatedly shrinking the range of the $m^v$ values, or to produce some evidence that this is not possible. A simplest version is the following:

\begin{algorithm}
\caption{\textsc{RepeatedPumping$(\epsilon)$}}
\label{algo1}
\begin{algorithmic}[1]
\STATE Initialize $h:=0$, and $x_h:=0\in\RR^V$. \label{(R0)}
\STATE Set $m^+(h):=\max_{v\in V} m^v(x_h)$ and $m^-(h):=\min_{v\in V} m^v(x_h)$.\label{(R1)}
\STATE If $m^+(h)-m^-(h)\leq \epsilon$ then STOP.\label{(R2)}
\STATE $x_{h+1}:=$\textsc{Pump}$(x_h,V)$; $h:=h+1$.\label{(R3)}
\STATE Goto step \ref{(R1)}. \label{(R4)}
 \end{algorithmic}
\end{algorithm}

Note that by our above analysis, \textsc{RepeatedPumping}  either returns a potential transformation for which all $m^v$, $v\in V$ values are within an $\epsilon$-band, or returns the sets $I$ and $F$ as in Lemma \ref{l2} with arbitrary large potential differences from the other positions. In the next section we use a modification of these procedures for stochastic games, and show that those large potential differences can be used to prove that the game is not $\epsilon$-ergodic.

\section{Application of Pumping for Stochastic Games}
\label{sg}
%As we argued in Section \ref{basic} when the game is ergodic, there exists a potential transformation after which all local matrix game values at the positions will be equal to the corresponding (global) game values, and the locally optimal strategies of players 1 and 2 will form globally optimal strategies for them. A similar statement can be made for the existence of $\epsilon$-ergodic solutions. 

We show in this section how to use \textsc{RepeatedPumping} to find potential transformations verifying $\epsilon$-ergodicity, or proving that the game is not $\epsilon$-ergodic, thus establishing a proof of Theorem \ref{t-main}. Towards this end, we shall give some necessary and sufficient conditions for
$\epsilon$-non-ergodicity, and consider a modified version of the pumping algorithm described in the previous section which will provide a constructive proof for the above theorem.

%We shall need to introduce additional notation and several technical lemmas.
%Given a potential $x\in \RR^V$, let us define
%\begin{equation}\label{e99}
%a_{k\ell}^{v}(x) ~=~ \sum_{u\in V}p_{k\ell}^{vu}(r_{k\ell}^{vu}+x^v-x^u)
%\end{equation}
%for all positions $v,u\in V$ and actions $k\in K^v$ and $\ell\in L^v$. For all positions $v\in V$ let us denote by $A^v(x)$ the $K^v\times L^v$ matrix formed by these entries, and define $m^v(x)$ as the value of the two-person zero sum matrix game with payoff matrix $A^v(x)$.

Let us first observe that the local value function of stochastic games satisfies the properties required to run the pumping algorithm described in the previous section.

\begin{lemma}\label{l3}
For every subset $S\subseteq V$ and $\delta\geq 0$ and for all $v\in V$ we have
\begin{equation}\label{e999}
\begin{array}{rcl@{\text{~~~~if~~}}l}
m^v(x)&\geq ~m^v(x-\delta e_S) &\geq ~m^v(x) - \delta\max_{k,\ell}\sum_{u\not\in S}p_{k\ell}^{vu}&v\in S,\\*[2mm]
m^v(x)&\leq ~m^v(x-\delta e_S) &\leq ~m^v(x) + \delta\max_{k,\ell}\sum_{u\in S}p_{k\ell}^{vu}&v\not\in S.\\*[2mm]
\end{array}
\end{equation}
Furthermore, the value functions $m^v(x)$ for $v\in V$ satisfy properties (i), (ii) and (iii) stated in Section~\ref{pump}.
\end{lemma}
\proof
According to \eqref{e99} we must have for all $\delta\geq 0$ that $A^v(x)\geq A^v(x-\delta e_S)$ for all $v\in S$ and $A^v(x)\leq A^v(x-\delta e_S)$ for all $v\not\in S$ proving properties (i) and (ii) (Indeed, $A^v(x-\delta e_S)=A^v(x)-\delta(E^v-\sum_{u\in S}P^{vu})$ for $v\in S$ and $A^v(x-\delta e_S)=A^v(x)+\delta\sum_{u\in S}P^{vu}$ for $v\not\in S$, where $E^v$ is the $|K^v|\times|L^v|$-matrix of all ones. Since the operator $\Val(B)$ is monotone increasing in $B$, inequalities \raf{e999} follow). Property (iii) follows directly from \raf{e999}. \qed

The above lemma implies that procedures \textsc{Pump} and \textsc{RepeatedPumping} could, in principle, be used to find a potential transformation yielding an $\epsilon$-ergodic solution. It does not offer, however, a way to discover $\epsilon$-non-ergodicity.
Towards this end, we need to find some sufficient and algorithmically achievable conditions for $\epsilon$-non-ergodicity.

Let us first analyze ($0$-)non-ergodicity of stochastic games (in stationary strategies).

\begin{lemma}\label{l01}
A stochastic game is non-ergodic if and only if it is $\epsilon$-non-ergodic for some positive $\epsilon$.
\end{lemma}

\proof
A stochastic game is non-ergodic by definition if there exists a threshold $\sigma$, positions $v,u\in V$,
and stationary strategies $\alpha$ and $\beta$ for the players, such that no matter what other strategy $\beta'$ player 2 chooses the Markov chain resulting by fixing $(\alpha,\beta')$ has a value $>\sigma$ when using initial position $v_0=v$ (guaranteeing for player 1 more than $\sigma$ from $v$), and the Markov chain obtained by fixing $(\alpha',\beta)$ has a value $<\sigma$ when using initial position $v_0=u$ (guaranteeing for player 2 less than $\sigma$ from $u$). Since strategies $\alpha'$ and $\beta'$ are chosen from a compact space, the above implies that there are $\sigma'>\sigma>\sigma''$ such that $\alpha$ guarantees for player 1 at least $\sigma'$ from the initial position $v$, and $\beta$ guarantees for player 2 at most $\sigma''$ from initial position $u$. Hence the game is $\epsilon$-non-ergodic for any $\epsilon ~<~ \sigma'-\sigma''$.\qed
\begin{lemma}\label{l02}
A stochastic game $\Gamma$ is $\epsilon$-non-ergodic if there exist disjoint non-empty subsets of the positions $I,F\subseteq V$, reals $a,b$ with $b-a\geq\epsilon$, stationary strategies $\alpha^v$, 
$v\in I$, for player 1, and $\beta^u$, $u\in F$, for player 2, and a vector of potentials $x\in \mathbb{R}^V$, such that
\begin{itemize}
\item[\emph{(N1)}]
$\alpha^v_k p_{k\ell}^{vu}=0$ for all $v\in I$, $u\not\in I$, $k\in K^v$ and $\ell\in L^v$,

\item[\emph{(N2)}]
$\beta^u_\ell p_{k\ell}^{uw}=0$ for all $u\in F$, $w\not\in F$, $\ell\in L^u$ and $k\in K^u$, and

\item[\emph{(N3)}] for all $v\in I$ and $u\in F$:
\[
\min_{\widetilde\beta^v\in\Delta(L^v)} (\alpha^v)^TA^v(x)\widetilde\beta^v ~\ge~ b ~~~\text{ and }~~~  \max_{\widetilde\alpha^u\in\Delta(K^u)} (\widetilde\alpha^u)^T A^u(x)\beta^u  ~<~a.
\]
\end{itemize}
\end{lemma}
\proof
Let us note that (N1) and (N3) imply that for all strategies $\beta'\in\Minstrat(\Gamma)$ of player 2, the pair of strategies $(\bar\alpha,\beta')$, where $\bar\alpha^v:=\alpha^v$ for $v\in I$ and $\bar\alpha^v\in\Delta(K^v)$ is chosen arbitrarily for $v\not\in I$, results in a Markov chain in which subset $I$ induces one or more absorbing sets (that is,
% no arc is leaving $I$ with a positive probability
$p_{\bar\alpha\beta'}^{vu}=0$), and in which all positions have values at least $b$. Analogously, (N2) and (N3) imply that $F$ will always induce an absorbing set with values less than $a$, if we fix any pair of strategies $(\alpha',\bar \beta)$, where $\alpha'$ is any strategy in $\Maxstrat(\Gamma)$, $\bar\beta^v:=\beta^v$ for $v\in F$ and $\bar\beta^v\in\Delta(L^v)$ is chosen arbitrarily, for $v\not \in F$. Hence choosing any positions $v\in I$ and $u\in F$ and strategies $\bar\alpha$ and $\bar\beta$ provides a witness for the $\epsilon$-nonergodicity of $\Gamma$.  (Here, we use the well-known fact \cite{MO70} that, to each player's stationary strategy, there is a best response of the opponent which is also stationary.)  
\qed

\smallskip

Let us introduce a notation for denoting upper bounds on the entries of the matrices, more precisely on the part of these entries which do not depend on negative potential differences. Specifically, define
\begin{equation}\label{e55}
\begin{array}{rl}
\widetilde{a}_{k\ell}^v(x)&=\displaystyle\sum_{u\in V}p_{k\ell}^{vu}r_{k\ell}^{vu} ~+~ \sum_{u\in V,~ x^u\leq x^v} p_{k\ell}^{vu}(x^v-x^u)\\*[5mm]
\widetilde{b}_{k\ell}^v(x)&=\displaystyle m^+(x)-\sum_{u\in V}p_{k\ell}^{vu}r_{k\ell}^{vu} ~-~ \sum_{u\in V,~x^u\geq x^v} p_{k\ell}^{vu}(x^v-x^u)
\end{array}
\end{equation}
where, as before, $m^+(x):=\max_{v} m^v(x)$, ~$m^-(x):=\min_{v} m^v(x)$. Define further

\begin{equation}\label{e78}
\begin{array}{rll}
R^v(x)&~=~\displaystyle\max_{k\in K^v, \ell\in L^v} \left(\widetilde{a}_{k\ell}^v(x)\right)&~~~\text{ if } m^v(x)\geq \frac{m^+(x)+m^-(x)}{2},\\*[2mm]
R^v(x)&~=~\displaystyle\max_{k\in K^v, \ell\in L^v} \left(\widetilde{b}_{k\ell}^v(x)\right)&~~~\text{ otherwise}.\\*[2mm]
\end{array}
\end{equation}

Note that $$m^+(x)-\widetilde{b}_{k\ell}^v(x)\le a_{k\ell}^v(x)\le\widetilde{a}_{k\ell}^v(x)~~~\text{ for all $v\in V,~k\in K^v,~\ell\in L^v$ and $x\in\RR^V,$}$$
which implies 
\begin{equation}\label{e11}
\begin{array}{rll}
m^v(x)&~\le~ R^v(x)&~~~\text{ if } m^v(x)\geq \frac{m^+(x)+m^-(x)}{2},\\*[2mm]
m^v(x)&~\ge~m^+(x)-R^v(x)&~~~\text{ otherwise}.\\*[2mm]
\end{array}
~~~\text{ for all $v\in V$ and $x\in\RR^V,$}
\end{equation}
With this notation we can state a more constructive version of Lemma \ref{l02}.

\begin{lemma}\label{l03}
A stochastic game $\Gamma$ satisfying \raf{e77} is $\epsilon$-non-ergodic if there exist disjoint non-empty subsets $I,F\subseteq V$, a vector of potentials $x\in \mathbb{R}^V$, and reals $a',b'\in[0,m^+(x)]$ with $b'-a'\geq 3\epsilon$, $a'<\frac{m^+(x)+m^-(x)}{2}$, $b'\ge\frac{m^+(x)+m^-(x)}{2}$,such that
\begin{itemize}
\item[\emph{(N4)}] $m^v(x)\ge b'$ for all $v\in I$, and $m^u(x)< a'$ for all $u\in F$;
\item[\emph{(N5)}] $x^u-x^v\geq |L^v|WR^v(x)^2/\epsilon$ for all $u\not\in I$, and $v\in I$;
\item[\emph{(N6)}] $x^u-x^v\geq |K^v|WR^v(x)^2/\epsilon$ for all $u\in F$, and $v\not\in F$.
\end{itemize}
\end{lemma}
\proof
We first show that (N4)-(N5) imply the existence of strategies $\alpha^v$, for $v\in I$, satisfying (N1) and (N3). We shall then observe that a similar argument can be applied to (N4) and (N6) to show the existence of strategies $\beta^u$, for $u\in F$, such that those satisfy (N2) and (N3). Consequently, our claim will follow by Lemma \ref{l02}.

Let us now fix a position $v\in I$ and denote respectively by $\bar\alpha^v$ and $\bar\beta^v$ the optimal strategies of players with respect to the payoff matrix $A^v(x)$. Denote further by $\widehat{\beta}^v=\frac{1}{|L^v|}(1,1,\ldots,1)$ the uniform strategy for player 2, and set $\bar K^v=\{k\in K^v\mid \sum_{u\not\in I} \sum_{\ell\in L^v}p_{k\ell}^{vu}=0\}$.

Let us then note that we have
\[
\left(A^v(x)\widehat{\beta}^v\right)_k\leq \left\{\begin{array}{ll}
R^v(x)&\text{ if } k\in \bar K^v,\\
R^v(x)-\frac{R^v(x)^2}{\epsilon}& \text{ otherwise,}
\end{array}\right.
\]
since at least one of the entries of (N5) has at least $\frac{W}{|L^v|}$ as a coefficient in rows which are not in $\bar K^v$.

%Since we had nonnegative rewards originally by \eqref{e77}, we must have 
Note that $b'>0$ implies by \raf{e11} that $R^v(x)>0$. Thus by the optimality of $\bar\alpha$ and by the above inequalities we have
\[
0<b'\leq m^v(x)\leq \bar\alpha^v A^v(x)\widehat{\beta}^v \leq R^v(x) -\left(\sum_{k\not\in \bar K^v}\bar\alpha_k^v\right)\frac{R^v(x)^2}{\epsilon}
\]
implying that
$
\sum_{k\not\in \bar K^v}\bar\alpha_k^v < \frac{\epsilon}{R^v(x)}.
$
Since by (N4) we have $0<a'$, inequalities $\epsilon<a'+3\epsilon\le b'< m^v(x)\leq R^v(x)$ follow, and hence $\frac{3\epsilon}{R^v(x)}<1$ must hold, implying that the set $\bar K^v$ is not empty. Let us then denote by $\widetilde{\alpha}^v$ the truncated strategy defined by
\[
\widetilde{\alpha}_k^v=\left\{\begin{array}{l@{~~~\text{ if }~~}l}
\displaystyle\frac{\bar\alpha_k^v}{\sum_{k\in \bar K^v}\bar\alpha_k^v}&k\in \bar K^v,\\*[3mm]
0&k\not\in\bar K^v.
\end{array}\right.
\]
With this we have for any $\widetilde\beta^v\in\Delta(L^v)$
\[
\begin{array}{rl}
b'\leq m^v(x)&\leq (\bar\alpha^v A^v(x)\widetilde\beta^v\\
&=\displaystyle\left(\widetilde{\alpha}^vA^v(x)\widetilde\beta^v\right)\left(\sum_{k\in \bar K^v}\bar\alpha_k^v\right) + \sum_{k\not\in\bar K^v}\bar\alpha_k^v\left(\sum_{\ell\in L^v}a_{k\ell}^{v}(x)\widetilde\beta^v_\ell\right) \\
&\displaystyle\leq \left(\widetilde{\alpha}^vA^v(x)\widetilde\beta^v\right) + \left(\sum_{k\not\in \bar K^v}\bar\alpha_k^v\right)R^v(x)\\
&< \left(\widetilde{\alpha}^vA^v(x)\widetilde\beta^v\right) +\epsilon.
\end{array}
\]

Let us then define $\alpha^v=\widetilde{\alpha}^v$ and repeat the same for all $v\in I$. Then, these strategies satisfy (N1) and (N3) with $b=b'-\epsilon$.

Let us next note that by adding a constant to a matrix game it changes its value with exactly the same constant. Furthermore, multiplying all entries by $-1$ and transposing it, changes its value by a factor of $-1$, interchanges the roles of row and column players, but leaves otherwise optimal strategies still optimal. Thus, we can repeat the above arguments for the matrices $B^u(x)=m^+(x)E^u-A^u(x)^T$, where $E$ is the $|L^u|\times|K^u|$-matrix of all ones, and obtain the same way strategies $\beta^u$, $u\in F$ satisfying (N2) and (N3) with $a=a'+\epsilon$.
This completes the proof of the lemma.\qed

\medskip

To create a finite algorithm to find sets $I$ and $F$ and potentials satisfying (N4)-(N6) we need to do some modifications in our procedures.

First, we allow a more flexible partitioning of the $m$-range by allowing the $m$-range boundaries to be passed as parameters and replacing line~\ref{(P1)} in procedure \textsc{Pump} by
\begin{itemize}
\item[2:] Set $\delta:=(m^+ -m^-)/4$.
\end{itemize}
 
Next, Let us replace in procedure \textsc{Pump}, line~\ref{(P4)} by the following lines, where $\epsilon>0$ is a prespecified parameter, and
call the new procedure with these modifications \textsc{ModifiedPump}$(\epsilon,x,S,m_-,m_+)$:
\begin{itemize}
\item[7a:] Otherwise set $P_\tau:=\{v\in S\mid m^v(x_\tau)\geq m^-+2\delta\}$ and compute
\[
\begin{array}{rl@{\text{ ~~~~if~~ }}l}
R_\tau^v&~:=~\displaystyle\max_{k\in K^v, \ell\in L^v} \left(\widetilde{a}_{k\ell}^v(x_\tau)\right)&v\in P_\tau,\\*[2mm]
R_\tau^v&~:=~\displaystyle\max_{k\in K^v, \ell\in L^v} \left(\widetilde{b}_{k\ell}^v(x_\tau)\right)&v\not\in P_\tau,\\*[2mm]
\end{array}
\]
where $\widetilde{a}$ and $\widetilde{b}$ are defined by \eqref{e55}.
\item[7b:] Create an auxiliary directed graph $G=(V,E)$ on vertex set $V$ such that $(v,u)\in E$ iff
\[
\begin{array}{r@{~~~\text{ if }~~}l}
x_\tau^u-x_\tau^v< \frac{|L^v|W\left(R_\tau^v\right)^2}{\epsilon}& v\in P_\tau,\\*[3mm]
x_\tau^v-x_\tau^u< \frac{|K^v|W\left(R_\tau^v\right)^2}{\epsilon}& v\not\in P_\tau.\\
\end{array}
\]
\item[7c:] Find subsets $I_\tau$ and $F_\tau$ of $V$ such that $T_\tau\subseteq I_\tau\subseteq P_\tau$, $B_\tau\subseteq F_\tau\subseteq V\setminus P_\tau$, and no arcs are leaving these sets in $G$ (this can be done by a finding the strong components of $G$, or by the method described int he proof of Theorem~\ref{t-main}).
\item[7d:] if such sets are found STOP and output these sets, otherwise continue with step \ref{(P5)}.
\end{itemize}

Before starting to analyze this modified pumping algorithm, let us observe that we have for all iterations
\begin{equation}\label{e33}
m^-<m^-+\frac{\epsilon}{2}< \frac{m^-+m^+}{2}\leq m^v(x_\tau) \leq R_\tau^v ~~~\text{ for all }~~~ v\in P_\tau
\end{equation}
as long as $m^+-m^->\epsilon$.

\begin{lemma}\label{l4}
Procedure \textsc{ModifiedPump}$(\epsilon,x,S)$ terminates in a finite number of steps.
\end{lemma}

\proof
Let us observe that by Lemma \ref{l2} procedure \textsc{Pump} would either terminate with $T_\tau=B_\tau=\emptyset$ for some finite $\tau\ge \tau_0$, or there exist sets $I=I_\tau$ and $F=F_\tau$ satisfying conditions (b) and (c) of the lemma, for $\Delta=NWQ^2/\epsilon$, where $N=\max\{\max\{|K^v|,|L^v|\}:~v\in I\cup F\}$, and $Q=\max\{R^v_{\tau(\Delta)}:~v\in I\cup F\}$.  
Thus, in the latter case, \textsc{ModifiedPump} will indeed find some sets $I_\tau$ and $F_\tau$, and hence terminate for some finite $\tau$.
%(Note that $R_\tau^v\leq R_{\tau(\Delta)}^v$ for all $v\in I\cup F$ and $\tau\ge\tau(\Delta)$, since the potentials of all vertices in $I\cup F$ change in the same way, for $\tau\ge\tau(\Delta)$, while $m^+(x_\tau)$ can only go down. It follows that $I_\tau=I$ and $F_\tau=F$ for all iterations $\tau\geq \tau(\Delta)$).
\qed

\begin{lemma}\label{l5}
Procedure \textsc{ModifiedPump}$(\epsilon,x,V)$ either shrinks the $m$-range by a factor of $3/4$ or outputs potentials $x=x_\tau$ and sets $I=I_\tau$ and $F=F_\tau$ which satisfy conditions (N4)-(N6) with $a'<b'$.
\end{lemma}

\proof
When the procedure terminates without shrinking the $m$-range, then it outputs sets $I=I_\tau$ and $F=F_\tau$ such that in the auxiliary graph $G$ there are no arcs leaving these sets. Since $I\subseteq P_\tau$ and $F\subseteq V\setminus P_\tau$, condition (N4) holds with $a'=\max_{v\not\in P_\tau}m^v(x_\tau)<b'=(m^++m^-)/2$. Furthermore, the lack of leaving arcs in $G$ implies that for all $(v,u)$, $v\in I$ and $u\not\in I$ and also for all $(u,v)$ with $u\in F$ and $v\not\in F$ we must have the reverse inequalities in (7b), implying that conditions (N5) and (N6) hold.\qed

%\bigskip

Let us observe that the bounds and strategies obtained by Lemmas \ref{l4} and \ref{l5} do not necessarily imply the $\epsilon$-non-ergodicity of the game since those positions in $I_\tau$ and $F_\tau$ may not have enough separation in $m$-values (i.e. the condition $b'-a'\ge 3\epsilon$ in Lemma~\ref{l03} is not satisfied).
To fix this we need to make one more use of the pumping algorithm, as described in the \textsc{ModifiedRepeatedPumping} procedure below.
After each range-shrinking in this algorithm, we use a routine called \textsc{ReducePotential$(\Gamma,x,m_-,m_+)$} which takes the current potential vector $x$ and range $[m_-,m_+]$ and produces another potential vector $y$ such that $\|y\|_\infty\le 2^{\poly(n,N,\eta)}$. We need to this because, as the algorithm proceeds, the potentials, and hence the transformed rewards, might grow doubly-exponentially high.  

The potential reduction can be done as follows. We write the following quadratic program in the variables $x\in\RR^V$, $\alpha=(\alpha^v~|~v\in V)]\in\Maxstrat(\Gamma)$, and $\beta=(\beta^v~|~v\in V)]\in\Minstrat(\Gamma)$:
 
\begin{align}
\alpha^vA^v(x')&\ge m_-\cdot\be,&A^v(x')\beta^v&\le m_+\cdot\be,\label{e1.1}\\
\alpha^v\be&=1,&\be\beta^v&=1,\nonumber\\
\alpha^v&\ge 0,& \beta^v&\ge 0,\nonumber
\end{align}
for all $v\in V$, where $\be$ denotes the vector of all ones of appropriate dimension. This is a quadratic system of at most $6N$ (in)equalities on at most $(2N+1)n$ variables. Moreover the system is feasible since the original potential vector $x$ satisfies it. Thus, a rational approximation to the solution to within an additive accuracy of $\delta$ can be computed,using quantifier elimination algorithms, in time $\poly(\eta,N^{O(nN)},\log\frac{1}{\delta})$; see \cite{BPR96,GV88,R92}. Note that the resulting solution will satisfy \raf{e1.1} but within the approximate range $[m_--\delta,m_++\delta]$. By choosing $\delta$ sufficiently smaller than the desired accuracy $\epsilon$, we can ignore the effect of such approximation.  
   
\begin{algorithm}
\caption{\textsc{ModifiedRepeatedPumping$(\epsilon)$}}
\label{algo3}
\begin{algorithmic}[1]
\STATE Initialize $h:=0$, and $x_h:=0\in\RR^V$.\label{(M0)}
\STATE Set $m^+(h):=\max_{v\in V} m^v(x_h)$ and $m^-(h):=\min_{v\in V} m^v(x_h)$.\label{(M1)}
\IF{$m^+(h)-m^-(h)\leq 24\epsilon$}\label{(M2)}
  \RETURN $x_h$.
\ENDIF
\STATE $x_{h+1}:=$\textsc{ModifiedPump}$(\epsilon,x_h,V,m_-,m_+)$ and let $F_\tau,I_\tau,T_\tau,B_\tau,P_\tau$ be the sets obtained from \textsc{ModifiedPump}.\label{(M3)}
\IF {$T_\tau=\emptyset$ or $B_\tau=\emptyset$}\label{(M4)}
  \STATE $x_{h+1}:=$\textsc{ReducePotential$(\Gamma,x_\tau,m_-(h),m_+(h))$}
  \STATE Set $h:=h+1$ and Goto step \ref{(M1)}
\ENDIF
\STATE Otherwise set $F=F_\tau$ and $I=I_\tau$.\label{(M5)}
%\IF{$\min_{v\in I_\tau}m^v(x_\tau)\ge \frac{9}{16}m_++\frac{7}{16}m_-$} \label{(M2)}
%  \STATE Goto step \ref{(M9)}
%\ENDIF
\STATE $x_{h+1}:=$\textsc{ModifiedPump}$(\epsilon,x_{h},I_\tau,m_-,m_+)$ and let $T_\tau,B_\tau$ be the sets obtained from this call of \textsc{ModifiedPump}.\label{(M6)}
\IF {$T_\tau=\emptyset$} \label{(M7)}
  \STATE $x_{h+1}:=$\textsc{ReducePotential$(\Gamma,x_\tau,m_-(h),m_+(h))$}
  \STATE Set $h:=h+1$ and Goto step \ref{(M1)}.
\ENDIF
\IF {$B_\tau=\emptyset$}\label{(M8)}
  \STATE Goto step \ref{(M9)}
\ENDIF
\STATE Otherwise, update $I:=I_\tau$.
\RETURN $x_{h+1}$ and the sets $I$ and $F$.\label{(M9)}
 \end{algorithmic}
\end{algorithm}

\begin{lemma}\label{t1}
\textsc{ModifiedRepeatedPumping}$(\epsilon)$ terminates in a finite number $
h\leq \log \frac{R}{24\epsilon}/\log \frac78,
$
of iterations, and either provides a potential transformation proving that the game is $24\epsilon$-ergodic, 
or outputs two nonempty subsets $I$ and $F$ and strategies $\alpha^v$, $v\in I$, for player 1 and $\beta^v$, $v\in F$, for player 2 such that conditions (N4), (N5) and (N6) hold with $b',a'$ satisfying the condition in Lemma~\ref{l03}.
\end{lemma}

\proof
Let us note that if $T_\tau=\emptyset$ after the second \textsc{ModifiedPump} call, then the range of the $m$-values has shrunk by a factor of $\frac78$ (at least), while if this happens in the first stage the $m$-range has shrunk by a factor of $3/4$.

On the other hand if the $m$-range is not shrinking, and we have $B_\tau=\emptyset$ after the second call of \textsc{ModifiedPump}, then we would also have $m^v(x_\tau)\geq\frac58m^++\frac38m^-=b'$ for all $v\in I$, while $m^u(x_\tau)< (m^++m^-)/2=a'$ for all $u\in F$, and hence (N4)-(N6) hold with these $a'$ and $b'$ values.
Since the $m$-range has not shrunk, we must have $m^+-m^->24\epsilon$, and hence $b'-a'=\frac18(m_+-m_+)>3\epsilon$ follows. (Note that, since in the second stage we pump only positions in $I_\tau$, the potentials of these positions may go down, while those of the positions outside $\cI_\tau$ remain unchanged, and hence condition (N5) remains satisfied.)  

Finally, if the $m$-range is not shrinking, and the second call returns a new set $I_\tau$, then all $m$-values of this set are at least $\frac34m^++\frac14m^->\frac58m^++\frac38m^-=b'$, and with the same set $F$ we can conclude again that conditions (N4)-(N6) hold.
\qed

%\bigskip

To complete the proof of Theorem \ref{t-main}, we need to analyze the time complexity of the above procedure, in particular, bounding the number of pumping steps performed in \textsc{ModifiedPump}.

Let us note that as long as $m^+-m^->24\epsilon$ we pump the upper half $P_\tau$ by exactly $\delta\geq 6\epsilon$.
Let $\cP_\tau(v)$ (resp., $\cN_\tau(v)$) denote the number of iterations, among the first $\tau$, in which position $v$ was pumped, that is, $v\in P_\tau$ (resp., not pumped, that is, $v\not\in P_\tau$).
 
Let us next sort the positions $v\in V$ such that we have
\[
x_\tau^{v_1} \leq x_\tau^{v_2} \leq \cdots \leq x_\tau^{v_n},
\]
and write $\Delta_j=x_\tau^{v_{j+1}}-x_\tau^{v_j}$ for $j=1,2,...,n-1$. Note that $\cP_\tau(v_1)=\tau$ and $\cN_\tau(v_n)=\tau$.

%Let us denote by $i \le i_\tau$ the largest index such that $v_i\in I_\tau\subseteq P_\tau$. 
Let $i_\tau$ be the largest index in $\{1,2,\ldots,n\}$, such that $v_{i_\tau}\in P_\tau$.
Then, by \eqref{e55} we have for $i=0,1,2,\ldots,i_\tau-1$ that
\begin{equation}\label{e90}
0\le \widetilde a_{k\ell}^{v_{i+1}}(x_\tau)\leq R+\sum_{j=1}^i\Delta_j,
\end{equation}
where the sum over the empty sum is zero by definition. Similarly, for $i=i_\tau+1,\ldots,n$, we have 
\begin{equation}\label{e93}
-R\le \widetilde b_{k\ell}^{v_{i}}(x_\tau)\leq R+\sum_{j=1}^{n-i}\Delta_{n-j}.
\end{equation}
From \raf{e90} and \raf{e93}, it follows that
\begin{equation}\label{e91}
|R_\tau^{v_{i+1}}|\leq\left\{
\begin{array}{lll}
& R+\sum_{j=1}^i\Delta_i,&\text{ for $i=0,1,2,\ldots,i_\tau-1$}\\
 & R+\sum_{j=1}^{n-i-1}\Delta_{n-j},&\text{ for $i=i_\tau,i_\tau+1,\ldots,n-1$.}
\end{array}
\right.
\end{equation}
Let $\widetilde i_\tau$ be the smallest index $i$ such that
\begin{equation}\label{e96}
\Delta_i> \frac{NW(R+\sum_{j=1}^{i-1}\Delta_j)^2}{\epsilon},
\end{equation}
and let $\widehat i_\tau$ be the largest index $i\le n-1$ such that
\begin{equation}\label{e97}
\Delta_i> \frac{NW(R+\sum_{j=1}^{n-i-1}\Delta_{n-j})^2}{\epsilon}.
\end{equation}
From the definition of $\widetilde i_\tau$, we know that 
\[
\Delta_i\le \frac{NW(R+\sum_{j=1}^{i-1}\Delta_j)^2}{\epsilon}, \text{ for all $i=1,\ldots,\widetilde i_\tau-1$.}
\]
Solving this recurrence, we get
\begin{equation}\label{e95}
x_\tau^{v_{\widetilde i_\tau}}-x_\tau^{v_1}=\sum_{i=1}^{\widetilde i_\tau-1}\Delta_i \leq  \left(\frac{(\widetilde i_\tau-1) NWR}{\epsilon}\right)^{2^{\widetilde i_\tau-1}-1} (\widetilde i_\tau-1)^2 R\le\left(\frac{n NWR}{\epsilon}\right)^{2^{n}-1} n^2R.
\end{equation}
Similarly, the definition of $\widehat i_\tau$ gives 
\[
\Delta_i\le \frac{NW(R+\sum_{j=1}^{n-i-1}\Delta_{n-i})^2}{\epsilon}, \text { for all $i=\widehat i_\tau+1,\ldots,n-1$,}
\]
from which follows
\begin{equation}\label{e98}
x_\tau^{v_n}-x_\tau^{v_{\widehat i_\tau+1}}\le\left(\frac{n NWR}{\epsilon}\right)^{2^{n}-1} n^2R.
\end{equation}
Note that if $\widetilde i_\tau\le i_\tau$ then \raf{e91} implies that taking $I_\tau=\{v_1,\ldots, v_{\widetilde i_\tau}\}$ would satisfy condition (N5) and guarantee that $I_\tau\subseteq P_\tau$. 

Indeed, for all $i\le\widetilde i_\tau$ and $u\not \in I_\tau$, we have
\[
x_\tau^u-x_\tau^{v_i}\ge\Delta_{\widetilde i_{\tau}}> \frac{NW(R+\sum_{j=1}^{i-1}\Delta_j)^2}{\epsilon}\ge \frac{|L^{v_i}|W\left(R^{v_i}(x_\tau)\right)^2}{\epsilon}.
\]

Similarly, having $\widehat i_\tau \ge i_{\tau}$ guarantees that taking $F_\tau=\{v_{\widehat i_\tau+1},\ldots,v_n\}$ would satisfy (N6) and $F_\tau\cap P_\tau=\emptyset$.

, since for all $i\ge\widehat i_\tau+1$ and $u\not \in F_\tau$, we have
\[
x_\tau^{v_i}-x_\tau^u\ge\Delta_{\widehat i_{\tau}}> \frac{NW(R+\sum_{j=1}^{n-i-1}\Delta_{n-j})^2}{\epsilon}\ge \frac{|K^{v_i}|W\left(R^{v_i}(x_\tau)\right)^2}{\epsilon}. 
\]

On the other hand, if $\widetilde i_\tau\ge i_\tau+1$, then \raf{e95} implies that $v_{i_\tau+1}$ was always pumped except for at most $\kappa(R):=\left(\frac{n NWR}{\epsilon}\right)^{2^{n}-1} \frac{n^2R}{\delta}$ iterations, that is, $\cN_\tau(v_{i_\tau+1})\le \kappa(R)$. Also, since $v_{i_\tau+1}\not\in P_\tau$, then at time $\tau$, $v_{i_\tau+1}$ is not pumped. Similarly, if  $\widehat i_\tau < i_{\tau}$, then \raf{e98} 
implies that $v_{i_\tau}$ was never pumped except for at most $\kappa(R)$ iterations, that is, $\cP_\tau(v_{i_\tau})\le \kappa(R)$, while it is pumped at time $\tau$. Since we have at most $n$ candidates for each of $v_{i_{\tau}}$ and $v_{i_{\tau}+1}$, it follows that after $\tau=2n\kappa(R)+1$, neither of these events ($\widetilde i_\tau\ge i_\tau+1$ and $\widehat i_\tau < i_{\tau}$) can happen, which by our earlier observations implies that the algorithm constructs the sets $I_\tau$ and $F_\tau$.
We can conclude that \textsc{ModifiedPump}$(\epsilon,x,V)$ must terminate in at most $2n\kappa(R)+1$ iterations, either producing $m^+-m^-\leq 24\epsilon$ or outputting the subsets $I_\tau$ and $F_\tau$ proving $\epsilon$-non-ergodicity.

One can similarly bound the running time for the second call of \textsc{ModifiedPump} (line~\ref{(M6)}), and the running time for each iteration of \textsc{ModifiedRepeatedPumping}$(\epsilon)$ (but with $R$ replaced by $2^{\poly(n,N,\eta)}$).

It remains now to bound the running time for the second call of \textsc{ModifiedPump} (line~\ref{(M6)}), and the running time for each iteration of \textsc{ModifiedRepeatedPumping}$(\epsilon)$. We can repeat essentially the same analysis as above, assuming that we modify the rewards with the potential vector obtained up to this point in time. Since, by the above argument, the maximum potential difference between any vertices before at the time $\tau$, when we make the second call to \textsc{ModifiedPump} is at most $\delta(2n\kappa(R)+1)$, it follows that the maximum absolute value of the transformed rewards at time $\tau$ is $r^{vu}_{k\ell}(x_\tau)\le R_2:= R+\delta(2n\kappa(R)+1)$ (note that the non-negativity of the rewards was only needed to bound $m_-\ge 0$ initially). It follows by the same argument as above that the second call \textsc{ModifiedPump} terminates in time $2n\kappa(R_2)+1=\left(\frac{n NWR}{\epsilon}\right)^{O(2^{2n})}$.
%, and the new transformed rewards are bounded in absolute value by $R_2+\delta(2n\kappa(R_2)+1)$.  

After shrinking the $m$-range, we apply potential reductions which guarantees that the bit length of each entry in potential vector is bounded by a polynomial in the original bit length $\eta$. It follows that the new transformed rewards will have absolute value bounded by $R_3=2^{\poly(n,N,\eta)}$.  We repeat the same  argument for the different phases of \textsc{ModifiedRepeatedPumping}$(\epsilon)$ to arrive at the running time claimed in Theorem~\ref{t-main}. 

 This completes the proof of the theorem.\qed

%\section{Closing Remarks}
%The so-called {\it switching controller} stochastic games were introduced by Filar \cite{Fil81}. In this special case, the set of positions is partitioned into two: $V=V_1\cup V_2$, where it is assumed that at each position only one of the two players controls the transition probabilities, i.e., if $V\in V_1$ (resp., $V\in V_2$), then $p^{vu}_{k\ell}=p_{vu}^k$ (resp., $p^{vu}_{k\ell}=p_{vu}^{\ell}$). Interestingly, in this case, it was shown in \cite{Fil81} that a saddle point exists in stationary strategies, and that it is satisfies the {\it order filed property}, i.e., there exist rational equilibrium strategies and values, if all the rewards and transition probabilities are rational. Let $W'$ denote the least common denominator of the rational probabilities $p^{vu}_{k\ell}$, and $k=|S|$ be the number of positions $v\in S\subseteq V$
%for which both $|L^v|>1$ and $|K^v|>1$ (i.e., positions controlled by both players). Using techniques similar to the ones in \cite{IPCO-2010}, we can show that the running time of the algorithm of Theorem \ref{t-main} can be improved to  $\left(\frac{NW'nR}{\epsilon}\right)^{O(k)}\poly(\eta)$, which is pseudo polynomial for constant $k$. Furthermore, with an accuracy of $\epsilon\approx\left(\frac{1}{nNW'}\right)^{O(kh)}$, where $h=\sum_{v\in S}|L^v||K^v|$, we can get an exact solution in time
%$\left(NW'nR\right)^{O(kh)}\poly(\eta)$.

\newcommand{\etalchar}[1]{$^{#1}$}

\end{document}